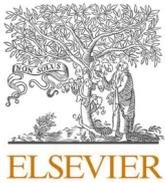
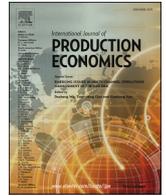

# Pivoting B2B platform business models: From platform experimentation to multi-platform integration to ecosystem envelopment

Clara Filosa [a,*], Marin Jovanovic [b], Lara Agostini [a], Anna Nosella [a]

[a] *Department of Management and Engineering, University of Padova, Stradella S. Nicola, 3, 36100, Vicenza, Italy*
[b] *Department of Operations Management, Copenhagen Business School, Solbjerg Plads 3, 2000, Frederiksberg, Denmark*

A B S T R A C T

The landscape of digital servitization in the manufacturing sector is evolving, marked by a strategic shift from traditional product-centric to platform business models (BMs). Manufacturing firms often employ a blend of approaches to develop business-to-business (B2B) platforms, leading to significant reconfigurations in their BMs. However, they frequently encounter failures in their B2B platform development initiatives, leading them to abandon initial efforts and pivot to alternative platform strategies. Therefore, this study, through an in-depth case study of a manufacturer in the energy sector, articulates a three-phase pivoting framework for B2B platform BMs, including platform development and platform strategy. Initially, the manufacturer focused on asset-based product sales supplemented by asset maintenance services and followed an emergent platformization strategy characterized by the rise of multiple, independent B2B platforms catering to diverse functions. Next, focusing on the imposed customer journey strategy, the firm shifted towards a strategic multi-platform integration into an all-encompassing platform supported by artificial intelligence (AI), signaling a maturation of the platform BM to combine a wide range of services into an energy-performance-based contract. Finally, the last step of the firm's platform BM evolution consisted of a deliberate platform strategy open to external stakeholders and enveloping its data-driven offerings within a broader platform ecosystem. This article advances B2B platform BMs and digital servitization literature, highlighting the efficacy of a progressive approach and strategic pivoting.

## 1. Introduction

The phenomenon of digital transformation is emerging as a new opportunity to stay competitive in the current economic scenario, marked by globalization, heightened competition, and frequent and substantial technological changes (e.g., AI) (Hanelt et al., 2021). In such a context, manufacturers are under constant pressure to innovate and adapt their business models (BMs) to the most cutting-edge trends (Cenamor et al., 2017; Jovanovic et al., 2022b; Sjödin et al., 2020), increasingly embracing the new digital servitization paradigm (Kohtamäki et al., 2019). Digital servitization refers to the creation of new or improvement of existing digital services, made smart by exploiting the potentialities of digital technologies for constantly collecting field data and monitoring product status, usage, and performance (Paschou et al., 2020). This new concept combines the potential of digitalization and servitization, through which manufacturers lay the foundations for new strategic opportunities, fundamentally altering their way of competing (Shen et al., 2023). Specifically, this dual focus enables companies to adopt both a customer- and service-centric lens using digital technologies and emphasizes customers' experience and engagement (Solem et al., 2022; Taylor et al., 2020; Vendrell-Herrero et al., 2017). Indeed, digital technologies are leveraged to gain knowledge about how customers use products, thus making manufacturers able to provide personalized solutions that fulfill specific demands and needs (Rymaszewska et al., 2017). By expanding and refining the scope of their new digital solution offerings, companies achieve closer and deeper engagement with customers, establishing long-term relationships (Dalenogare et al., 2023). Digital servitization, in turn, implies higher performance for firms in terms of competitiveness, economic sustainability, customer loyalty, and profitability (Markfort et al., 2021; Martín-Peña et al., 2019).

However, while extant research has focused on the economic aspects of digital servitization, several empirical studies highlight that traditional manufacturers struggle to concretely and successfully embrace this new economic paradigm and reap the benefits derived therefrom (Weking et al., 2020). Specifically, despite the growth of research in literature, there is still a dearth of theory development on the digital servitization journey that leverages B2B platform development (Cenamor et al., 2017; Jovanovic et al., 2022b; Saadatmand et al., 2019; Wei et al., 2019). B2B platform development has proven to be an effective tool for enacting the digital servitization transformation (Eloranta et al., 2021). Adopting a platform approach entails an extension of relationships with external stakeholders (i.e., customers, suppliers, and partners) from a dyadic involvement to a multilateral ecosystem environment (Sjödin et al., 2020), and the development of an advanced platform architecture that enables aggregating various data types and formats and advancing AI platform service deployment (Sjödin et al., 2023). A manufacturing firm may choose to develop one or multiple types of B2B platforms, such as transaction, innovation, or hybrid platforms (Gawer, 2014). Yet, they all come with distinct managerial challenges. For example, developing transaction-driven B2B marketplaces (Lanzolla and Frankort, 2016; Wamba et al., 2008) may bring a different set of network effects in comparison to traditional B2C platforms (Thomas et al., 2024). Similarly, developing an industry-wide platform may require a distinct platform architecture (e.g., blockchain) and careful platform governance since the platform owner can also be a competitor (Jovanovic et al., 2022a). Finally, developing an innovation-driven or hybrid B2B platform ecosystem (Jovanovic et al., 2022b) with AI applications can be challenging to monetize, given the difficulty in inducing and developing complementors (Saadatmand et al., 2019), especially if the provider already has highly advanced performance-based contracts in place (Visnjic et al., 2017).

There is currently a lack of knowledge regarding how manufacturers strategize about and manage B2B platform development (Volberda et al., 2021), considering the supply-side and the demand-side of the platform market (Bonina et al., 2021). Previous studies on the topic highlight the necessity of a holistic transformation to successfully embrace digital transformation and the creation of a platform BM (Volberda et al., 2021), showing the different configurations a platform-based BM can assume (Ritala and Jovanovic, 2024). Furthermore, different studies analyze the role of the platform orchestrator (Shen et al., 2024), especially in the industrial markets (Van Dyck et al., 2024). However, detailed knowledge on how to practically implement this new meta-organizational form in terms of new routines, structures, and processes, is still limited (Kretschmer et al., 2022). It has not been thoroughly studied, specifically, how platform owners handle strategic challenges and opportunities (Ghazawneh and Henfridsson, 2011; Staykova, 2018), when switching from a product-centric to a platform ecosystem approach (Stonig et al., 2022; Van Dyck et al., 2024). For instance, emergent, imposed, and deliberate approaches (Mintzberg and Waters, 1985) to platform strategy have been less studied in the literature, specifically when advanced AI services are deployed (Bawack et al., 2022; Fosso Wamba et al., 2022).

Next, B2B platform strategies are also directing new B2B platform BMs that alter the processes of value creation, value delivery and value capture mechanisms (Anderson et al., 2022; Ritala and Jovanovic, 2024; Tian et al., 2022). More importantly, with the promising rise of AI functionalities, there is a need to better understand how B2B platform BMs evolve, and how this leads to the redesign of manufacturers' activity systems as well as the orchestration mechanisms with external actors (e.g., customers and partners), with the goal of co-creating value (Ritala and Jovanovic, 2024; Shen et al., 2024). Therefore, the link between platform strategic response and the characteristics of B2B platform BMs remains understudied, representing an important area to explore (Van Dyck et al., 2024). Specifically, how manufacturers pivot from one B2B platform BM to another (Burnell et al., 2023; Shepherd et al., 2023; Snihur and Eisenhardt, 2022), resulting in platform portfolios (Tarzijan and Snihur, 2024) and multi-platform integration (Schreieck et al., 2024), is scarcely explored in the literature.

Finally, little is known about how manufacturers pivot into platform ecosystems to generate value through complementor applications that support AI services (Clough and Wu, 2022) and envelop their data-driven offerings within broader platform ecosystems (Geurts and Cepa, 2023; Ritala et al., 2024).

The identified gaps highlight significant inadequacies in the existing literature, underscoring the need for theoretical development on key aspects of digital servitization to support and guide manufacturers in successfully navigating this transformative process. Therefore, this study aims to contribute to the discussion about the intersection of B2B platform development, B2B platform BMs, and associated platform strategies. Specifically, tackling the outlined gaps, we investigated the process of implementing advanced AI services. Accordingly, our study addresses the following research question:

*How do incumbent manufacturing firms embrace different platform strategies and pivot their B2B platform BMs to provide AI services?*

To address this question, we first reviewed the literature on digital servitization, AI, platform strategies, and B2B platform BM pivoting to build a solid theoretical background for our analysis. Next, we performed an in-depth qualitative single case study to describe the digital servitization process of an incumbent manufacturer in the energy sector. The analysis highlights the journey the case company has gone through in implementing AI services, closely examining the evolution of the company's platform strategies, value propositions, as well as the progression of platform development. We argue that the journey evolves over time through a progressive process, with phases that can be delineated. Specifically, the study theorizes on the evolution from platform experimentation to multi-platform integration to ecosystem envelopment, describing how the manufacturer adapts its platform strategy, platform development, and platform BM across the identified phases. Ultimately, this study proposes a pivoting framework that connects platform strategies and platform development, with platform BMs, showing how those elements and their intertwining evolve. By doing so, the article contributes to the literature by expanding the knowledge of digital servitization in relation to B2B platform development. This contributes to improving the discussion on the role of B2B platforms, specifically in the design and development of AI services. Moreover, the insights gathered through the in-depth case study allow us to understand how traditional manufacturers strategize about B2B platforms and pivot their platform BMs accordingly. The valuable understanding of those elements, along with the empirical knowledge derived from the case study investigation, support managers of traditional manufacturing firms in delineating the proper strategic planning and actions for a successful digital servitization transformation, thus also highlighting the managerial contribution of the article.

## 2. Theoretical background

### 2.1. Digital servitization and artificial intelligence

The new paradigm of digital servitization derives from the interplay of the two phenomena that are affecting manufacturing businesses today: digitalization and servitization. Digitalization represents the process of transformation of procedures, operations, and offerings within companies by leveraging the potential of new and advanced digital technologies to enable product and service improvements, enhanced efficiency and effectiveness (Gebauer et al., 2021; Matthyssens, 2019; Taylor et al., 2020; Tronvoll et al., 2020). On the other hand, servitization consists of enriching companies' value propositions by complementing the offering of physical goods and basic services with advanced services (Visnjic et al., 2017; Weking et al., 2020), thus providing customers with integrated, personalized solutions that able to satisfy specific customer needs (Frank et al., 2019b; Kohtamäki et al., 2019). The combination of digitalization and servitization leads to the innovative and advanced digital servitization concept, consisting of the transformation of processes, capabilities, and offerings to progressively create, deliver, and capture value through services based on enabling digital technologies (Shen et al., 2023). These digital technologies are leveraged to provide monitoring, optimization, and autonomous solutions, thus driving value for both providers and customers (Frank et al., 2019b; Paiola and Gebauer, 2020).

In this context, digitalization represents a driving force, and servitization is the output (Kohtamäki et al., 2022). A full-scale digital servitization is not possible without adequate data acquisition and analytics (Frank et al., 2019a; Kohtamäki et al., 2020). Digital technologies are indeed adopted to connect products, resources, and people in a smart way (Porter and Heppelmann, 2015; Raff et al., 2020), collecting and generating a high volume and variety of data (Hsuan et al., 2021). Specifically, data from a large pool of customers, installed bases, infrastructures, and systems are stored, analyzed, and elaborated into meaningful insights (Jovanovic et al., 2022b). While operations for data collection and analysis can also be performed by human intervention, which can take a long time, digital technologies enable easier coding and classification of unstructured data from products, turning it into useable and user-friendly information (Rezazade Mehrizi et al., 2023; Subeesh and Mehta, 2021). By aiding the human effort, still required to check the proper operation of the technologies and algorithms underpinning the data analysis (Rezazade Mehrizi et al., 2023), this facilitates the development of a tailored value proposition to meet customers' needs in the most responsive and reliable way (Hasselblatt et al., 2018; Iriarte et al., 2023). Digital servitization also enhances the decision-making process by providing constantly updated field data and analytics about critical product parameters and status, and exploiting knowledge throughout the asset lifecycle (Kohtamäki et al., 2022). For example, Internet of Things (IoT) sensors are used to accumulate massive data on physical products; cloud computing serves as data warehousing; big data and analytics render field insights into valuable information and predictive patterns that enable value-creating solutions (Grubic, 2018; Paiola and Gebauer, 2020; Suppatvech et al., 2019).

Despite the strategic significance of digital servitization mentioned above, the role of cutting-edge technologies such as AI has still not been appropriately investigated (Sjödin et al., 2021; Thomson et al., 2023). AI is defined as "*a system's ability to interpret external data correctly, to learn from such data, and to use those learnings to achieve specific goals and tasks through flexible adaptation*" (Kaplan and Haenlein, 2019). In practice, AI consists of algorithmic computer systems that may accomplish intricate tasks. These include data processing operations that enable the use of learning algorithms, the discovery of complex patterns in high-dimensional data, and the generation of output forecasts (Bauer et al., 2023; Jordan and Mitchell, 2015). Consequently, AI can be used in a wide range of scientific, business, and decision-making contexts (LeCun et al., 2015).

In the context of digital servitization, AI is exploited for its capability to accurately assess collected data, to elaborate on those inputs, to draw conclusions from them, and to apply those conclusions to particular activities and goals (Sjödin et al., 2021). Specifically, AI plays a crucial role in the evolution of cognitive cyber physical systems (CPS) by enabling autonomous decision-making processes where physical activities affect computations and vice versa (Radanliev et al., 2021). These characteristics of AI entail multiple functionalities (Hansen and Bøgh, 2021). Among others, descriptive analytics techniques are used by AI algorithms to summarize insights and data collected through IoT sensors, presenting them as user-friendly information (Berente et al., 2021). Moreover, machine learning can be exploited for diagnostic purposes where the analytics are able to identify and classify association or causal relationships among events, for example, detecting the reasons behind a machine malfunction (Hansen and Bøgh, 2021). Another case is the use of AI in offshore windmills, where AI can automatically cancel operations in case of high wind power to prevent damage. Hence, the integration of AI in CPS allows for the processing of high volume and variety of data (e.g., high search scope) and the solving of previously unexplored problems (Raisch and Fomina, 2024). This exemplifies AI prescriptive capability, which allows autonomous decision systems to help devices adjust to their environment (Hansen and Bøgh, 2021). Finally, AI generates knowledge and lays the groundwork for automating processes (Raisch and Krakowski, 2021) by highlighting patterns that constitute the basis for the creation of new products and services (Berente et al., 2021; Brynjolfsson and Mitchell, 2017). Therefore, by taking advantage of these AI features and functionalities, manufacturers can aim to offer advanced, innovative, and around-the-clock services like performance advisory, predictive maintenance, and autonomous solutions (Abou-Foul et al., 2023; Sjödin et al., 2021), which may also support circular economy (Sjödin et al., 2023). The early literature on autonomous solutions suggests that these services are highly context-dependent, making it essential to pilot smaller-scale trials to assess their feasibility before making full-scale investment decisions, a crucial step for implementing AI solutions in industrial settings (Thomson et al., 2023). In turn, AI solutions demand a high level of context understanding and entail significant investment costs, which may not be feasible if the solution is not replicable (Thomson et al., 2023). Therefore, strategic pivoting may accommodate these challenges by allowing firms to adjust their strategies based on initial trials and adapt their AI solutions to specific contexts.

*2.2. Digital servitization strategic pivoting and business model reconfiguration*

Digital servitization holds considerable significance for manufacturing firms, presenting an opportunity to differentiate themselves from competitors, cater to a broader spectrum of partners and customers, and tap into new market opportunities by reaching previously inaccessible markets and enhancing supply chain efficiency (Anderson et al., 2022; Cenamor et al., 2017; Jovanovic et al., 2022b; Shree et al., 2021; Tian et al., 2022). However, its practical implementation remains a formidable challenge for firms. This difficulty can be attributed to the inherent nature of digital servitization that, like any substantial organizational change, necessitates a strategic response (Autio, 2022; Van Dyck et al., 2024; Volberda et al., 2021) and disrupts the current BM (Martinez, 2022). This involves a comprehensive redefinition of the company's strategic objectives, encompassing the realignment of plans, activities, and targets to resonate with the newly adopted paradigm. As such, digital servitization represents a strategic pivot, defined as the transformation resulting from economic experimentation (Pillai et al., 2020; Gomes et al., 2021) and, hence, from the learning process that arises out of the introduction of a new technological or market paradigm (Grodal et al., 2015). As stated by Greenstein (2007, p.2), the deployment of the new strategic pivot "*leads to changes in business operations and organizational procedures that translate technological innovation into market value*", creating the opportunity for the introduction of a new product or service.

The concept of pivoting is extensively discussed in the entrepreneurship literature as a crucial mechanism during early-stage BM experimentation (Felin et al., 2020; Ghezzi and Cavallo, 2020). Pivoting involves strategic shifts to test new fundamental hypotheses about the product, strategy, and growth engine, which may entail significant changes to the BM, value proposition, target market, or technology (Shepherd and Gruber, 2021). Therefore, digital servitization pivoting requires a fundamental redesign of value creation, value delivery, and value capture mechanisms (Teece, 2010), demanding the formulation and implementation of novel managerial practices (Kohtamäki et al., 2022; Suppatvech et al., 2019). Yet, the BM reconfiguration seldom occurs in a singular, abrupt transformation; instead, it typically manifests as a gradual progression that invariably affects the BM dimensions (Schreieck et al., 2024). The economic experimentation initiated by digital servitization pivoting, in fact, is a learning process over time across multiple aspects, involving the whole firm's structure and its strategic and market position (Gomes et al., 2021; Pillai et al., 2020). Using the terminology provided by Grodal et al. (2015), a new opportunity like digital servitization disrupts the industry and individual firms' equilibrium, which is operationalized by the firm reconfiguring its fundamental BM dimensions: value creation, value capture, and value delivery (Shen et al., 2023; Rabetino et al., 2017).

First, value creation is the development of an offering that is able to satisfy market demands and for which customers recognize value and utility (Teece, 2010; Sjödin et al., 2020). With the drastic shift towards digital servitization, value is mainly created by exploiting data constantly collected through digital technologies (Kohtamäki et al., 2022) that are leveraged to provide customers with the right solution that responds to their needs (Taylor et al., 2020) through performance and usage monitoring and optimization. Second, value delivery consists of the processes, activities, and means used to provide the defined value proposition to customers, thus identifying the way the business reaches the market (Sjödin et al., 2020; Teece, 2010). Digital services are often provided with the aid of technologies through which the company establishes a direct relationship with the customers and optimizes internal activities: technological support, in fact, enables firms to monitor products and information flows, adjusting the supply of offering according to the backend activities and resources and frontend requirements (Sjödin et al., 2020; Dalenogare et al., 2023). Finally, value capture represents the mechanisms through which a company can profit from its offerings, defining the proper cost structure and revenue model (Linde et al., 2023). Focusing on the cost structure, by exploiting digital technologies, manufacturers are able to take track of customers' needs and prioritize interventions based on field data collected in real time. By doing so, providers can allocate resources and develop client and product knowledge that can be used to improve the company's offering more effectively (Abou-foul et al., 2021; Linde et al., 2023). Hence, beyond the initial investment, digital servitization may enable significant cost efficiencies. Revenue stream mechanisms are also subject to reformulation: the provision of services allows firms to establish long-term relationships with customers for the entire physical product lifecycle, substituting one-shot transactions that characterize the selling of physical goods, and adopt a different pricing logic, obtaining recurring revenues with monthly or yearly fees paid to ensure the correct functioning of assets through well-designed digital services (Kohtamäki et al., 2019).

Yet, many uncertainties still emerge about how digital servitization reconfigures BMs and how B2B platforms facilitate this transition (Frank et al., 2019a; Paiola and Gebauer, 2020), depending on the context in which firms operate, market characteristics, existing operations, and ways of doing business. This holds true especially for manufacturers whose B2B platforms are associated to the development of advanced services and solutions that profoundly alter their own value creation, value delivery, and value capture processes while also changing the activities of their customers (Iriarte et al., 2023). However, studies rarely adopt a process perspective in industrial contexts to support manufacturers in recognizing and overcoming these limitations (Jovanovic et al., 2022a, 2022b).

### 2.3. The role of B2B platforms and B2B platform strategies

The relevance of B2B platforms in digital servitization is becoming increasingly pronounced (Anderson et al., 2022; Jovanovic et al., 2022b; Stonig et al., 2022; Van Dyck et al., 2024). The B2B platform serves as a pivotal cyber-physical hub for information and data exchange (Colombo et al., 2017), laying the groundwork for the development of diverse digital services (Cenamor et al., 2017) including AI-powered autonomous solutions (Thomson et al., 2023). Adopting a platform approach not only facilitates, but also actively generates new market opportunities (Cennamo, 2021). First, B2B transaction platforms connect buyers and sellers, streamline the purchasing process and reduce traditional market friction, thereby creating a more efficient marketplace that encourages broader participation from diverse, often global, partners and customers (Fan et al., 2023; Lanzolla and Frankort, 2016; Nambisan et al., 2019). Second, B2B platforms facilitate ecosystem creation, enabling complementary firms to collaborate and collectively develop innovative solutions, opening up new opportunities and leading to the emergence of entirely new market segments (Foss et al., 2023; Jacobides et al., 2018, 2024; Jacobides, 2022). Third, B2B platforms enhance data transparency by prioritizing standardized data sharing and aggregation (Agrawal et al., 2019; Berente et al., 2021), which builds trust among participants, facilitates better decision making via effective governance mechanisms (Autio, 2022; Shen et al., 2024), as well as monetizing data (Ritala et al., 2024). Finally, as more participants join the B2B platform, its value increases due to network effects, attracting additional users and partners, further expanding the market, and creating a self-reinforcing cycle of growth (Chen et al., 2022; Rietveld and Schilling, 2021).

Therefore, on the one hand, the B2B platforms, both transaction-based and innovation-based (Gawer, 2014; Gawer and Cusumano, 2014), foster a more profound relationship between the firm and its customers and partners. This is achieved not only by analyzing transaction data or product data in the field, but also, and most importantly, by actively involving customers and partners in designing the most suitable digital services and solutions to meet their specific needs. Consequently, customers and partners become integral participants in the co-creation process of digital services and solutions (Dalenogare et al., 2023; Tian et al., 2022).

On the other hand, the data sharing mechanism on which the functioning of a B2B platform is based also raises challenges in terms of privacy, security, trust, ethics, data ownership, and governance (Agrawal et al., 2019; Berente et al., 2021). When a serious risk of important information being leaked or misused is perceived, with consequent damage to competitive advantage, stakeholders could withdraw from the platform (Berente et al., 2021; Jovanovic et al., 2022a). Hence, setting rules, not only for the operation of the algorithms that elaborate products data into meaningful information and predictions, but also for properly handling the governance mechanisms among the stakeholders involved, becomes fundamental (Agrawal et al., 2019). For example, ensuring that multiple users have access only to pertinent information for their duties and guaranteeing and boosting the necessary dialogue between front-end and back-end functions distributed across several actors, help in defining roles and distributing the needed activities for a successful commercialization of advanced digital services and solutions (Cenamor et al., 2017).

Given the potentialities, advantages, and issues posed by the introduction of a B2B platform, selecting an appropriate platform strategy becomes essential. The formation of a platform strategy can be analyzed according to the theoretical lens provided by Mintzberg and Waters (1985) who identify three distinct strategic paradigms: emergent, deliberate, and imposed strategies (Mintzberg and Waters, 1985). Emergent strategies are characterized by patterned actions that are not the result of explicit intentions set by senior management, while deliberate strategies are closely aligned with the organization's articulated intentions. In contrast, imposed strategies originate from external sources, particularly customer influence, which directly shapes the company actions. This last paradigm represents a hybrid form, where the realized strategy merges the firm's objectives with external customer pressures to ensure competitive success. These three paradigms may succeed each other, reflecting the learning process established by the experimentation of a strategic reorientation: using the terminology provided by Grodal et al. (2015), when a new opportunity as digital servitization disrupts the industry and individual firms' equilibrium, this creates a strategic reorientation, characterized by an initial divergence and a subsequent convergence. The initial disruption characterized by uncertainty and experiments proceeds through improvements and selection, taking into account internal adjustments and, most of all, market feedback, which ultimately translates into a mature and more structured development of the realized strategy (Grodal et al., 2015).

In the context of B2B platform pivoting, a firm strategically changes its focus from one paradigm to another (Schreieck et al., 2024). For example, this strategic reorientation often involves transitioning from a single-sided to a multi-sided platform, or vice versa, based on evolving customer and stakeholder needs, with the ultimate goal of enhancing scalability, leveraging network effects, and boosting overall value creation (McDonald and Eisenhardt, 2020). Moreover, studies suggest that B2B firms usually start off with the role of a product-service platformizer, building a one-sided innovation platform that supports their installed bases, and subsequently focus on unlocking the demand side through collaboration with key customers and partners, occupying the role of platform ecosystem orchestrator (Ritala and Jovanovic, 2024). However, understanding of how manufacturers strategize (Volberda et al., 2021) about B2B platform development remains understudied. In particular, research on how B2B manufacturers effectively integrate specific platform mechanisms into their BM is scant (Tian et al., 2022; Cenamor et al., 2017; Shepherd and Gruber, 2021), as is the assessment of the strategic impact of AI (Berente et al., 2021) for digital servitization, and the investigation of the convergence of platform strategy and operationalization into an innovative BM (Huikkola et al., 2022; Lafuente et al., 2023), thus calling for further insightful research.

## 3. Methodology

### 3.1. Research design and case selection

Given the exploratory nature of our study and the goal of illustrating the range of platform strategies, events, dynamics, and challenges the digital servitization transition presents to manufacturers, we adopted an inductive research method based on a comprehensive single qualitative case study. This approach is well suited for the study because it enables the investigation of new and transformational phenomena and their real-world applications (Yin, 2018). Specifically, by ensuring access to a large and insightful amount of information, the case study approach provides detailed and sound explanations for "how" and "why" questions, contributing to truly understanding the multifaceted phenomena of digital servitization transformation and the BM reconfiguration deriving therefrom (Chen et al., 2021; Paiola and Gebauer, 2020; Weking et al., 2020). To deeply comprehend the significant dynamics and complexities of the research topic, we adopted a process perspective, examining the case while things were happening, thus also boosting the results' validity, and reducing the possibility that informants would not accurately recall relevant events (Langley, 1999; Langley et al., 2013). Moreover, through the adoption of a process perspective, we were able to organize events and business transformations into distinct temporal spaces, which made it possible to comprehend the case story in great detail and recognize a progression over time (Decker, 2022). The use and combination of a single case study and a process perspective are considered appropriate in further advancing knowledge of novel, complex phenomena, as in our research case, and have been already applied in previous research (Chen, 2020; Decker, 2022; Tronvoll et al., 2020). In particular, the effort that a process approach and a real-time viewpoint entail validates our decision to focus on just one case study to thoroughly examine the phenomenon that is the subject of this article (Pettigrew, 1990) and to contextualize the nature of the events (Tronvoll et al., 2020).

For this purpose, we selected an insightful case company, based on the following selection criteria: (a) the firm is an incumbent B2B company operating in the manufacturing industry; (b) it is involved in digital servitization projects using B2B platforms that render AI services; (c) it serves as an example of a visible process of pivoting, allowing us to gather useful insights to address our research question; (d) it acknowledges the significance of the research and makes a commitment to it; (e) it offers consistent access to data and informed sources, being available to collaborate, which facilitates meetings and interviews (Mason, 2002). In line with the research design and criteria, we investigated the case of a manufacturing company (Alpha) operating in the energy sector.

### 3.2. Data collection

The research process progressed through multiple stages, beginning with data collection and continuing through review and validation during the analysis phase. During data collection, both primary and secondary data were gathered combining a pluralistic approach (Evers et al., 2017). Primary data were collected from July 2023 to April 2024 through semi-structured interviews and workshops, following both real-time and retrospective perspectives, covering any relevant information for the analysis of our research topic (Pettigrew, 1990). Interviews and workshops were conducted with key informants: executive managers and project managers actively engaged in developing and implementing the digital servitization project were selected as sources of information to ensure material validity. An overview of data collected from key informants is reported in Table 1 below.

For the first interview sessions, with the aid of an interview guide, open-ended questions were posed to the respondents. The development of the semi-structured questionnaire was driven by themes related to the digital servitization journey, platform development, platform BM reconfiguration, and expected outcomes. Examples of questions used to discuss the main topics of interest are: *What is your current value proposition and how has it changed over the years? What are the platform services you offer to your industrial clients? What is your approach to platform development? How has your platform strategy evolved throughout the digital servitization journey? How has platform usage evolved over time? How do you leverage platform and AI technologies? How do you exploit the platform for the implementation of AI service offerings? How did digital technologies change your platform BM? What are the value creation and value capture mechanisms you associate with platform services, and how did you define the proper ones? What role have customers played in your digital servitization journey? Are there external partners involved in digital services activities? If so, which ones and why? What are the main challenges you have experienced along the digital servitization journey? What is the evolution of the digital servitization journey in the near future?*.

To truly comprehend Alpha's transition to digital servitization, during the course of meetings and interviews, we also used a forward-looking approach, posing foresight questions and assessing the digital servitization trajectory for the near future (Sarpong et al., 2019; Sarpong and O'Regan, 2014), inquiring about the changes that are already planned and in the process of being implemented, but not yet entirely executed.

During the first meeting, a clarification about the digital servitization concept was provided by the researchers, to set a common knowledge base among the participants, although interviewees were knowledgeable about it. Researchers posed questions, opening up deep and frank conversations with the interviewees. Discussions with the respondents

**Table 1**
Overview of data collection.

| Respondents' position | Interview/workshop | Date | Duration | Topic |
| --- | --- | --- | --- | --- |
| Product lead | Interview | July 4, 2023 | 33 min | Transaction-driven B2B platform |
| Executive | Interview | Jul. 28, 2023 | 35 min | Preliminary meeting |
| Executive | Interview | Aug. 17, 2023 | 63 min | Presentation of DS project |
| Executive, Product lead, Manager | Workshop | Dec. 20, 2023 | 240 min | As-is digital customer channels and B2B platform usage and to-be digital servitization project roadmap |
| Executive | Follow-up meeting | Apr. 25, 2024 | 46 min | Clarification and validation of concepts, discussion on on-going business changes |

clarified the relevant themes and were useful for obtaining further complementary information.

As the collaboration with the selected company continued, more detailed information was shared with the researchers. An open discussion on these topics required a more in-depth debate, which led to the organization of a workshop, to dedicate the right time to each of the themes on the agenda and their proper analysis. Specifically, the workshop was dedicated to a deep understanding of the company's digital servitization journey, leveraging a platform approach, reconfiguring its BM, and adapting the corporate strategy. A follow-up meeting was finally organized to share and discuss preliminary results with the interviewees, to detect any incomplete, incongruent, or missing information and to validate the authors' conceptualization. In addition, the insights coming from the direct interaction with Alpha's key respondents were further enriched and validated through triangulation of analysis with company reports and website materials (Jick, 1979; Yin, 2018). Specifically, to increase the accuracy and robustness of the collected information, secondary data from the firm website, internal presentations, reports, and official documents were employed (Yin, 2018). These data also allowed us to comprehend the context in which the selected manufacturer operates and in which the digital servitization transformation has occurred (Decker, 2022).

### 3.3. Data analysis

Interviews and the workshop were recorded and transcribed to increase accuracy and guarantee the possibility of consulting valuable insights *ex-post*. Interview and workshop notes were carefully examined jointly by the researchers. First, a comprehensive overview of the selected case was obtained, identifying the main relevant aspects related to its digital servitization transition. For this purpose, we followed the methodology developed by Gioia et al. (2013) that allowed us to classify and cluster the themes pertinent to our study. First, an initial reading and coding of primary data from interviews were conducted by all the researchers to identify essential information. Second, after processing the list of key information, also with the support of secondary data, similarities and overlaps were identified, producing a list of first-ordered categories of empirical concepts, grouping quotes, notes, and events. These categories were then aggregated and linked to each other in a meaningful way, generating second-order conceptual themes. Finally, these were further consolidated and abstracted into aggregate dimensions. Combining the inductive approach with an abductive logic, in the last step of the analysis, we labeled the aggregate dimensions by connecting the findings from our empirical evidence with prior literature. Indeed, the concepts were not completely new, thus leading to a theory-driven coding. The result of the data analysis process is synthesized in Fig. 1 that shows the coding structure and highlights the three aggregate dimensions, namely "Platform strategy evolution", "Platform development evolution" and "Platform business model evolution". Results will be presented according to the phases of the digital servitization journey, where each phase will be characterized in terms of platform strategy evolution, platform development evolution, and platform BM evolution.

## 4. Findings

### 4.1. Research context

Alpha is a manufacturing firm in the energy sector, producing energy solutions for utility companies. The company's avant-garde spirit has allowed it to be recognized globally as a manufacturer of highest-quality energy solutions, able to take advantage of experience in the industry as well as of the technologies and solutions invented over time, always remaining at the forefront of the market. Alpha has developed an extensive portfolio of energy solutions, associated not only with the basic services of installation, repair and previous consultancy to design the most adaptive energy solutions for clients, but also with advanced on-demand maintenance services. Through this solid business strategy, Alpha has positioned itself as a leading manufacturer in the energy assets market.

Notwithstanding the leading position Alpha has been able to establish in the industry, the company must deal with the ever-changing context in which it operates. The intense global competition and the growing difficulty of maintaining a product-based differentiation strategy have changed the balance of the energy sector. By exploring and adopting digital technologies as enablers of advanced digital services and by expanding its value proposition to increasingly offer personalized solutions to its customers, the firm has been able to take advantage of these market challenges and stand out in the market. Specifically, considering and exploiting the twenty- to thirty-year lifecycle of the energy assets and the extreme relevance of energy-efficient solutions, the company has enriched its offering in recent years: in addition to the energy assets, Alpha provides customers with complete energy solutions, based on advanced digital services aimed at achieving energy efficiency and ensuring constant optimal performance of the assets. Indeed, in a context where downtimes, failures, anomalies, repair and maintenance represent major operational costs, digital solutions based on the effective real-time remote control of the customer equipment constitute a winning strategy.

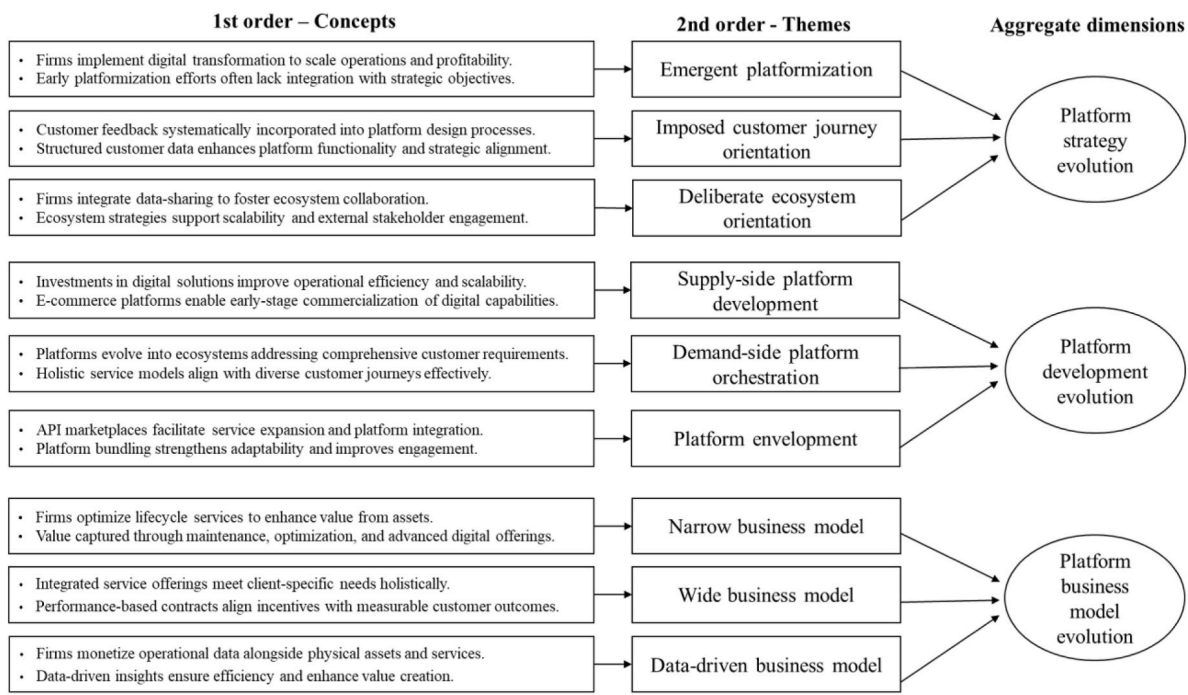

**Fig. 1.** Data structure.

The following paragraph presents this company transformation. Specifically, the analysis of the case leads to the division of this digital servitization transformation journey into different phases, providing a precise periodization of Alpha transition (Karsten, 2014). By examining and ranking the noteworthy occurrences, three primary phases of the transformation process are delineated, underlining an evolution that has recently resulted in a more mature and integrated approach to digital servitization, based on the learning experience developed over time. The three identified phases are referred to as platform experimentation, multi-platform integration, and ecosystem envelopment. Information related to key dynamics, mechanisms, and organization was codified to deeply understand what characterizes the firm and the features of the platform strategy and different BM components in every identified phase, as well as the evolution of the platform approach, in order to understand the main changes Alpha has experienced along the way, with the aim of successfully implementing digital servitization, from the beginning of the process until Alpha current configuration. The timeline of Alpha evolution from being a traditional manufacturer to achieving platform ecosystem envelopment is laid out in Fig. 2 below.

An in-depth description of the firm transformation process, starting from Phase 1, is provided in the following paragraphs. Specifically, we examine the evolution of the strategic plan adopted by Alpha throughout the identified phases using the theoretical lens provided by Mintzberg and Waters (1985). Moreover, we use the classification provided by Teece (2010) to identify the BM structure in use and the main changes along the digital servitization journey. Finally, we assess how the firm has leveraged platforms across the identified phases to enable the provision of advanced digital services, and we provide a description of the new offering.

The results shed light on how our case company used a platform approach to handle the shift to digital services, outlining the key changes and factors that affected it. The analysis of the results emerging from the case study allows us to answer the research question, unveiling the evolutionary process adopted by the selected manufacturer and how it reconfigured its strategy and BM and leveraged an increasingly integrated platform approach to support the digital servitization transformation.

### 4.2. The digital servitization journey

#### 4.2.1. Phase 1: Platform experimentation

*4.2.1.1. Platform strategy: Emergent platformization.* In Phase 1, Alpha embraced and invested in digitalization to push forward its value proposition. As described in a company's report, this included the creation of digital online channels, as well as the deployment of advanced digital technologies, such as AI, IoT, big data, and analytics, associated with physical components and designed to make it easier and more efficient to operate the energy assets portfolio through field data gathering:

> "We digitally transform our business to enable scalability and profitability".

Thus, by leveraging digital technologies, Alpha enriched its offering, shifting towards digital servitization. In its first attempt to achieve this new strategic goal, Alpha allocated resources to digital servitization projects separately, developing different initiatives. The executive stated:

> "We tried to digitalize more to see how we can act. This includes exploring new avenues in our service business and integrating advanced technologies into our processes".

Specifically, the company developed multiple digital channels as communication, informative and commercial contact points with customers, creating a multi-platform approach to service customers in different ways. This platformization approach enabled Alpha to establish a strong digital infrastructure that allowed the provision of multiple services, which represented a first emergent deployment of the digital servitization paradigm by experimenting with a portfolio of projects.

*4.2.1.2. Platform development: Supply-side platform development.* Alpha embraced the digital servitization paradigm using multiple digital channels as supportive tools. In particular, in Phase 1, as a B2B manufacturer, the firm internally developed digital platforms for different purposes. The executive of Alpha explained:

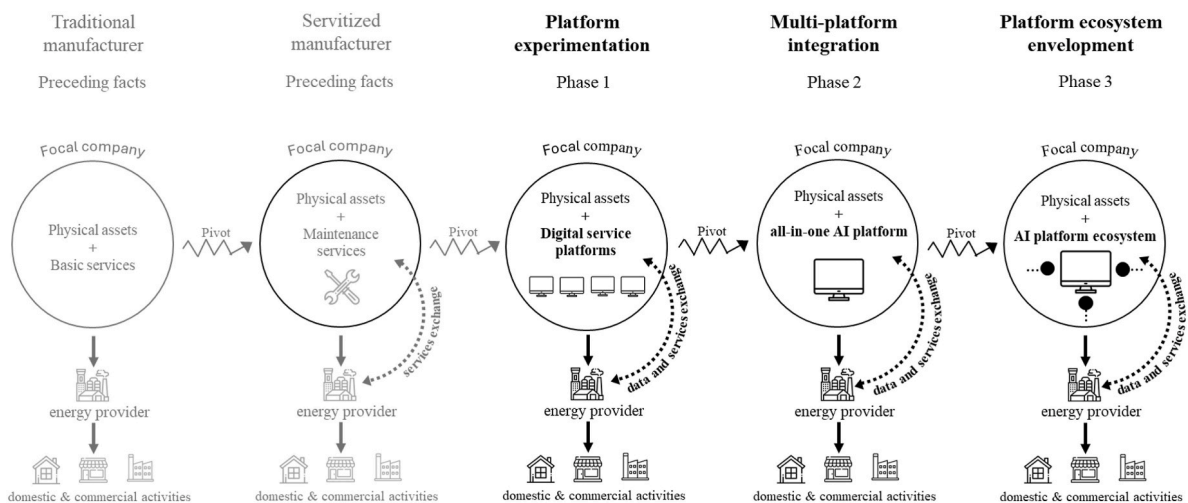

**Fig. 2.** Alpha's digital servitization journey.

*"We invested in digital solutions and e-commerce to streamline our operations, striving to be more than just a manufacturer, but a complete service provider in the energy sector".*

First of all, Alpha digitalized the already implemented maintenance service activity, designing and deploying an asset-based data platform. It functioned as a reporting dashboard for sharing relevant energy fleet and asset data as well as performance reports transparently, demonstrating to customers their fleet operational status and, eventually, the need to intervene. Hence, the primary innovation brought about by the development and deployment of the asset-based data platform was the access to diagnostic results and insights at the fleet, asset, and individual asset component levels, as described in Alpha's website:

*"It is a web-based dashboard available as an application that gathers inspection data across sites, [assets] and [components] into one collective overview. The dashboard is interactive and easy to use, allowing you to explore inspection results across geographies, [asset] models and [component] variants".*

Access to the platform was offered for free to all customers that had a commercial relationship with Alpha, providing access to technical and contractual information (i.e., invoices, technical documentation, safety alerts). Additional premium services were instead provided upon the subscription of a contract and, consequently, the payment of a fee. In this scenario, two alternatives were offered to the client. In the first option, the customer signed a performance-based contract in which Alpha committed itself to guarantee assets up-time, certified through monthly reports built thanks to field data collected and processed by advanced digital technologies. Related to this scenario, the executive stated:

*"We have some customers for whom we might not want to share the inspection information; we use it internally as part of the contract to maintain and optimize the [component] functioning".*

This kind of contract implied that Alpha was responsible for the correct functioning of the customer's fleet, thus providing services as repair, maintenance, and fleet optimization, ensuring customers their fleet operational efficiency. The second option consisted of offering customers access to all the data collected from their fleet, displaying them in a transparent way through the online platform, so that customers can use that information to train their own technicians and to perform self-maintenance activities for the correct management of the assets. In this specific case, the executive explained:

*"We share inspection with specific customers on their specific asset base and this might trigger a dialogue and a commercial opportunity with them based on the recommendation on that. So the customer can choose to commission us prepared to intervene and recover the product, paying an additional price. Or they can choose to do it themselves".*

Second, Alpha digitalized the purchasing process with a spare parts e-commerce platform with a worldwide reach: a B2B platform dedicated to the aftersales market in the energy sector. Customers benefited from an online procurement and quotation channel developed as a shop-like website. Automating and streamlining the transactional purchasing process enabled customers to place orders of spare parts and consumables, driving efficiency both for Alpha and its clients. A product lead explained:

*"The e-commerce platform initially started as an initiative within Alpha out of an awareness of the lack of consolidation in the market around making all of the types of spare parts that are needed across [assets] models more easily available to the customers that need them".*

Third, the company also launched an aftermarket platform for the entire energy sector with the aim to effectively match suppliers and purchasers of parts and services from various energy technologies, aiming to turn the disjointed aftermarket into a linked and streamlined online marketplace. Recognizing the importance of neutrality in such a marketplace, the product lead highlighted the need for the platform to be perceived as independent. In this regard, the product lead elaborated:

*"For a B2B marketplace to really work, it really has to be perceived as a more independent party that wouldn't position other sellers or other manufacturers in a disadvantageous position".*

Finally, the firm also acquired a smart data analytics platform for the energy sector, through which energy asset operators had access to best-in-class analytics products across energy sources, providing predictive guidelines for decision making, thanks to advanced AI data science. The characteristics of this platform are well described in Alpha's website:

*"Alpha offers customers a suite of best-in-class analytics products across energy sources that supports the digitalization of energy assets and energy systems. Giving easy access to portfolio-wide asset visualization, predictive maintenance, and power forecasting, the suite represents a critical business decision-making tool".*

The implementation of the above-described multiple digital platforms to approach customers and expand the value proposition testified Alpha's attempts to introduce a digital servitization strategy and demonstrated that this was a path-dependent, evolutionary process. The executive stated:

*"We started to commercialize the online spaces and learn what we should commercialize in the future … it's a slow process. We cannot just build, you know, without understanding problems and users".*

*4.2.1.3. Platform business model: Narrow business model.* Following and adopting the digital servitization paradigm through a platformization approach, Alpha modified its business model (BM). Looking at the value creation, the most evident change was the enrichment of the firm value proposition: next to the physical product and basic services, Alpha offered a variety of digital channels and thus access to multiple advanced services. The executive explained that, with the introduction of digital channels and related services:

*"We get into services around the [assets] and how we can optimize from planning to installations on till the end of operating".*

Next to the e-commerce and marketplace platform and the smart data analytics at market level, the greater innovative contribution of the digital servitization strategy for Alpha's customers was the offering of real-time information and insights at fleet, assets, and component level, guaranteeing ongoing performance of the assets thanks to the provision of the necessary services.

Hence, in Phase 1, separate service contracts represented the main company's output. In this new setting, the category of clients that Alpha addressed and served remained unchanged compared to the previous phase, as well as the direct approach it used, without intermediaries. However, the firm added supplementary channels of contact, sale, and communication that allowed constant and transparent interaction points between the firm and the customers. By doing so, a continuous online relationship between the customer and the firm replaced the "one-shot" interaction that characterized traditional manufacturers, leading to stronger trust and collaboration, intensified also by data and information sharing.

The development and deployment of the digital platforms also resulted in a reconfiguration of activities and capabilities, which was necessary to ensure the success of this innovation, as well as to guarantee alignment with the already existing business centered around physical assets. In particular, with the increased focus on digital servitization in Phase 1, a function dedicated to the digital servitization projects was established to deploy the whole set of activities required for offering the new value proposition. The product lead stated:

*"Some years back we created the development function, where we go in, design and mature new proposals".*

Specifically, as described in one company report:

*"Development, our newest business area, helps our customers grow their business […]. More than 100 employees across 15 countries secure land rights and permits, design sites, ensure grid connection and secure project offtake agreements to create quality projects".*

Furthermore, digital technologies became key assets Alpha had developed to successfully embark on this transformation. The executive explained:

*"IoT inspection data and photos and videos are collected by Alpha in a "component inspection repository" and there is an assessment of the [assets] status and some recommendations are elaborated. We use AI and analytics tools for the inspection of images and photos and for elaborating the recommended actions accordingly. Then, humans have to do the validation of the last decision. This information is provided and pulled out to the customers. Customers have access to information at the fleet level, the [asset] level and at the single [component] level in the dedicated platform, where they can also find and look at the photos and videos".*

Hence, thanks to this cutting-edge technological structure, Alpha could move towards a performance-based contract offering. When signing the service contracts, customers had two options: the first is receiving these insights through the platform, where they can view inspection data, indications of discovered damages with linked levels of severity, photographs proving the harm, and receive guidance on what actions to take, so that they can use those data for properly handling their fleets and organize self-service activities carried out by their own technicians; the second is that Alpha used field data to ensure the achievement of power generation promised to the client. In some cases, for big clients that had fleets in different sites around the world, these types of contracts could co-exist, as testified by the executive's words:

*"The same customer can have a site for the 100 [assets] in [country] for which we do a full-service activity and then the same customer can have a site in [country] where they do everything by themselves. So, the same customer can be both a partnership customer with a full scope contract and a more basic contract".*

Consequently, the value capture component of the BM also underwent a reconfiguration: beyond the income deriving from the sale of energy assets, the firm registered revenues for the provision of advanced digital services throughout the asset lifecycle; selling spare parts and separate service contracts became a significant source of income. Specifically, revenues came from selling field data to those customers that then implemented self-service activities for maintaining their fleets constantly in use; or, alternatively, customers paid for an energy power performance guaranteed by Alpha that was in charge of carrying out the needed digital services on the basis of the actual asset status, monitored in real time through advanced digital technologies. In both cases, Alpha received an additional recurrent revenue through the payment of yearly or monthly fees, depending on the agreement in place.

In this regard, the company distinguished clients and asset types. With long-standing customers, the company established twenty-year contracts for certain assets and five- or ten-year contracts for other assets, taking care of the entire management. With customers with whom it had recently established commercial relationships or who wanted to purchase data and not services, contracts were signed on an annual basis. In any case, Alpha ensured additional and recurring revenue for the entire asset lifecycle, thus guaranteeing itself a stronger economic sustainability.

*4.2.2. Phase 2: Multi-platform integration*

*4.2.2.1. Platform strategy: Imposed customer journey orientation.* Moving on, in Phase 2, Alpha embarked on a strategic initiative to develop a new operational commercial model, based on the learning experience of the previous phase that highlighted the need to involve customers. The executive explained:

*"There was a learning process, I don't know if it was conscious or more hind-sight. We need customers' involvement in the design process, in finding out what to build and how to build, in defining whether it works or not … we need customers' feedback on it: that is part of the learning process. We became more systematic in how we collect insights from customers now".*

Alpha recognized that the previous phase represented a primitive experimentation of the digital servitization strategy, and, as such, fallacies occurred. The most significant one is represented by the closure of the energy aftermarket platform. In this regard, the executive stated:

*"We are a manufacturing company and we are experts in energy solutions and we ended up with a very technical platform solution and we probably lost sight of what was really our value proposition for customers … Moreover, we missed some fundamental capabilities to make the platform work successfully and we also underestimated the maturity of the other actors in the field. Putting all these elements together, it was a complex situation. The customers' feedback was that the idea was good but somehow we came out with selling what we were already selling. We did not add value or something new. We become aware that customer is the starting point and that we have to be customer-centric".*

Drawing on the previous experience, Alpha became aware that a successful strategy required a customer-centric approach and decided to implement what customers were asking for. They applied what Mintzberg and Waters (1985) define as "imposed strategy": the external environment, specifically the clients, shapes the organization's course of action and determines what the business can accomplish. In the analyzed case, Alpha moved towards a strategic approach that guaranteed a better solution offering for the customers. Specifically, Phase 2 represented a consolidation interval with respect to the previous phase, based on a new commercial model predicated on the discontinuation, integration, and expansion of existing multiple digital platforms to establish a comprehensive platform ecosystem in the energy sector. The focus was on creating a consistent customer journey orientation and laying the foundation for the subsequent phase of expansion into an ecosystem. In this consolidation phase, the executive clarified:

*"We organized a number of different services and now we want to make them available in a connected, meaningful way. […] We need to get into it (talking about the digital solution they offer) with a totally different scale. […] We need to consolidate what is really the ambition for our digital customization across the company".*

*4.2.2.2. Platform development: Demand-side platform orchestration.* The strategy described above marked a shift towards a more cohesive and customer-centric digital servitization framework, aiming to enhance customer value through a demand-side platform orchestration of service delivery. On different occasions, the executive stated:

*"We are working on a new commercial model for a holistic view of the customer to provide solutions along the entire customer journey".*

And:

*"We said: ok, now we have these platforms and we need to consolidate to get a more coherent ecosystem. We need to move from the multichannel space and have a more coherent and a single-sign-on platform that is easier to access for the customer".*

Hence, the firm worked on an integrated platform, described in a company report as follows:

*"It is a one-stop entry point that connects the customers and Alpha for all the relevant needs and tasks, creates connected journeys across asset- and customer life-cycle, seamlessly connects fleet overview to relevant data and tools, and connects the customers to value creating offerings and recommendations".*

For example, whenever the integrated platform detected an anomaly in the asset functioning or status, Alpha would signal it through the platform, which would suggest the customer the right spare part or component needed to fix the problem, offering the possibility to directly buy it, and, at the same time, Alpha would send its expert technicians into the field to perform the service activity. The executive provided an additional example:

*"We want the customer to say: "ok, I am within Alpha's platform and these are the services available to me as a specific customer and I might be different from the customer next to me". So, there is a personalized offering: based on the collected field insights we know this customer needs a component to be changed in 6 months, so we send a notification to them to advise them and we recommend to substitute the component possibly in that specific date, because the forecasts are good enough and the spare part prices are low, so this date is the optimal point. This is just one example of how we connect things".*

*4.2.2.3. Platform business model: Wide business model.* The adoption of the integrated platform model signified a shift for Alpha towards fostering a solution that combines a wide array of services. This solution enabled customers to seamlessly access a variety of services ranging from spare part purchases to real-time data analytics and expert recommendations, all consolidated within a single platform.

Since the integrated platform model streamlined services for customers, the value creation dimension demanded continuous innovation and maintenance to keep up with evolving technological trends and customer expectations, posing potential challenges in sustaining long-term value creation.

Moreover, the consolidation of diverse services into a single platform required robust and scalable infrastructure, which could strain Alpha operational capabilities and resources, especially in managing increased complexity and ensuring consistent service delivery, as the product lead stated:

*"In the long term this creates a lot of transformation across the company … there is the need to change what has been done for 20 years in Alpha".*

The executive confirmed:

*"Having a platform landscape entails more value, but it also adds more complexity: services create value themselves but they introduce complications from a customer perspective. Services have to be developed on an individual customer basis, but, at the same time, Alpha needs to change the architecture and be smarter".*

Moreover, the internal configuration of activities had to be revised, adding complexity to the corporate structure for the sake of an improved customer experience, as the executive clearly explained*:*

*"Normally you have a marketing department, then you have a sales department and a service department and a development function and that's fine, but I want to put them all together because we believe that this generates opportunities for my company as well as for the customers".*

The challenge was even greater: despite the internal structure changed and there was a greater degree of complexity, which also led to higher costs, the value capture mechanism remained unchanged compared to the previous phase: customers purchased a multi-year service contract, with the difference that, in terms of value delivery, there was a facility for the customer to access a single platform having everything they needed at their fingertips.

*4.2.3. Phase 3: Ecosystem envelopment*

*4.2.3.1. Platform strategy: Deliberate ecosystem orientation.* To complete the integration, Alpha strategy included the involvement of external stakeholders and partners that may contribute to service offerings. To do so, it opened APIs (i.e., application programming interfaces) that are instrumental in achieving a fluid merger with the systems of customers and partners. This initiative not only bolstered the overall customer experience, but also broadened Alpha service capabilities, underscoring the inherent value of the ecosystem approach that can be exploited to build a seamless, industry leading customer experience. This change represented a deliberate shift towards an ecosystem orientation: a more open strategy that adopts a broader initiative that extends outside the firm boundaries, involving other industry stakeholders and actors. The executive stated:

*"We were then looking at the bigger ecosystem. […] We expose these services that we are in control of, but we are also part of an ecosystem where we need to connect our data to others and the other way around we need to connect others into our ecosystem".*

*4.2.3.2. Platform development: Platform envelopment.* The consolidation of functionalities into a unique platform performed in the previous phase opened up the possibility to further expand the digital infrastructure, creating a platform ecosystem, with different partners involved. Specifically, Alpha fed the platform with documentation and alerts; the technicians - both internal workforce and external partners - accessed it to deeply understand the identified problem, study the field data and allow a purposeful intervention; customers entered the platform for any kind of activity, recognizing it as a unique solution platform. In this phase, an ecosystem approach was applied thanks to a further enrichment of the platform features: Alpha opened APIs, giving customers and external partners the possibility to connect their own information systems with Alpha platform, thus supporting easy integration and building an industry leading customer experience. In this respect, the executive stated:

*"We're going to build an API marketplace and then we're going to give some of these services as an API as well. We will monetize some of these data to a large extent in the future. Basically, you can become a partner with Alpha API or you can browse applications and insights in the marketplace. Plenty of different applications even related to compliance, connectivity, fuel maintenance, market intelligence, optimization, and training, are available, and these apps can all be co-developed with the API".*

*4.2.3.3. Platform business model: Platform ecosystem business model.* Opening up the AI platform to external stakeholders entailed a series of business model changes. As the executive said, the value creation centered no longer on products and digital service contracts, but:

*"The core of our business is also energy efficiency (offering customers user-friendly insights and services to obtain the optimal energy output from the [assets]) … We offer what we call "power solutions" … not just the uptime, but we also guarantee the output".*

Looking at the value capture mechanism, the ecosystem model, while presenting expanded revenue opportunities, simultaneously introduced complexities in revenue management. A key challenge laid in devising effective strategies for charging for data sold to the external partners. Specifically, according to Alpha executive, next to assets and service contracts, the firm "*monetizes field data*" by linking key information access to payments. This would balance the costs incurred from integrating advanced technologies and multiple partners and the imperative to maintain profitability, all while navigating the diverse demands of customers and the fluctuating dynamics of the market. The complexity

*C. Filosa et al.*            *International Journal of Production Economics 280 (2025) 109466*of aligning the pricing model of these services with the pre-determined terms of performance-based contracts added an additional layer of complexity to Alpha business model in this emerging ecosystem.

In essence, while the ecosystem model offered significant advantages in terms of customer engagement and service integration, it also presented unique challenges to Alpha business model, particularly in the realms of sustaining value creation, managing complex service delivery, and capturing value in an increasingly interconnected and dynamic market environment.

The synthesis of the results is reported in Fig. 3 that provides a representation of the three phases identified during this study, detailing the platform strategy, the platform development and the business model at the different phases of the evolution.

## 5. Discussion and conclusion

### 5.1. Theoretical and managerial implications

This study provides an in-depth exploration of how manufacturing companies are strategically and structurally pivoting their B2B platform BMs by developing B2B platforms to offer advanced AI services. While these complex value offerings are essential to meet the growing and diverse needs of customers, they are not without their challenges. Successfully deploying a digital servitization strategy and implementing advanced digital solutions require a deep understanding of how to effectively structure these offerings and the overarching platform BMs. In light of this, our study makes several significant theoretical and managerial contributions, addressing the intricacies and dynamics of this transformative process in the B2B manufacturing sector.

First, this study argues that manufacturing firms undergo a strategic reorientation and pivoting from an emergent approach to an imposed one, ultimately applying a deliberate strategy in B2B platform development (Kohtamäki et al., 2019, 2022). The succession of these phases reflects the process of strategic platform pivoting (McDonald and Eisenhardt, 2020; Shepherd and Gruber, 2021), defined as the transformation resulting from a platform experimentation (Gomes et al., 2021; Pillai et al., 2020) and, hence, from the learning process that arises out of the introduction of a new technological or market paradigm (Grodal et al., 2015) such as digital servitization (Gomes et al., 2021). The results of our case study contribute insights into how companies transition from initial, *ad-hoc* supply-side separate platform initiatives to more structured and integrated demand-side approaches in platform development, ultimately building a platform ecosystem (Bonina et al., 2021; Shi et al., 2024). Moreover, the study also underscores that companies may initially experiment with multiple platforms to maximize learning and gather diverse insights, before later converging on a more deliberate and uniform platform strategy in the form of platform ecosystem envelopment (Schreieck et al., 2024). This progression from exploratory multiple platforms to an integrated platform ecosystem approach reflects an adaptive learning process, guiding firms towards more efficient and cohesive digital servitization strategies (Sjödin et al., 2021).

Second, this study contributes to the understanding of AI integration in B2B platforms within the manufacturing context. It highlights how AI-driven platforms are leveraged for processing real-time data, enabling the creation of personalized solutions for customers, thereby transforming traditional manufacturing models into platform BM (Fosso Wamba et al., 2022; Ritala and Jovanovic, 2024).

Third, the study shows the comprehensive reconfiguration of platform BMs in manufacturing firms as they increasingly embrace digital servitization. It provides a nuanced perspective on how these firms adapt across all BM dimensions, including value creation, value delivery, and value capture, in response to digital advancements (Sjödin et al., 2021). Moreover, Alpha investigation reveals that the BM reconfiguration does not occur at once, rather, it unfolds progressively (Jovanovic et al., 2022b).

Furthermore, this study explores the development of a unified platform ecosystem (Van Dyck et al., 2024) and ecosystem envelopment in the manufacturing sector (Eisenmann et al., 2011; Geurts and Cepa, 2023). It contributes to the understanding of how integrating various services into a single, cohesive platform ecosystem can enhance customer experience and operational efficiency, while also highlighting the complexities and revenue management challenges inherent in this approach (Jovanovic et al., 2022b), including selling and monetizing data in B2B markets (Ritala et al., 2024). This emerges clearly in the last stage of Alpha's transition, where the convergence of AI capabilities and platform ecosystem not only provides a single customer entry-point to various services and solutions, but it also enables a seamless integration and orchestration among the several stakeholders and partners (e.g., the focal company, customers, technology developers).

Finally, the study enters into the debate on the digital servitization course and, more specifically, on the discussion on whether it commences from digitalization or servitization. The majority of studies tends and provides evidence to support the first hypothesis, showing the pivotal role of digital technologies for integrating a customer-centric approach based on the sales of personalized solutions (Harrmann et al., 2023; Schulz et al., 2023) or identifying servitization as a mediator in the relationship between the deployment of digital technologies and the company's financial performance (Abou-Foul et al., 2021; Davies et al., 2023; Yang et al., 2023). In contrast, the deep and detailed case study reported in this article demonstrates that even the second route is possible. In doing so, the article supports and endorses the results by Vendrell-Herrero et al. (2024): the case of Alpha illustrates how a manufacturer may first integrate services into its value proposition and, once servitized, use digital technologies to scale those services. In this specific scenario, the platform flawlessly serves the scalability pathway. Through the integration of digital technology inside a servitized BM, a manufacturer can effectively optimize services, deliver them remotely, and customize them to meet the unique needs of individual clients.

| | Platform strategy evolution | Platform development evolution | Platform business model evolution |
|---|---|---|---|
| **Phase 1: Platform experimentation** | *Emergent platformization*<br>Resource allocation to digital servitization projects that are initiated separately | *Supply-side platform development*<br>Internal development of four separate platforms: Assets data platform, e-commerce platform, energy aftermarket platform, energy data analytics platform | *Narrow business model*<br>Asset-based sales and fleet optimization services separately |
| **Phase 2: Multi-platform integration** | *Imposed customer journey orientation*<br>Resource allocation to digital servitization projects that are unified into a unique consistent concept of strategy, requested by customers | *Demand-side platform orchestration*<br>Single unified platform | *Wide business model*<br>All-in-one performance-based contracts |
| **Phase 3: Platform ecosystem envelopment** | *Deliberate ecosystem orientation*<br>Resource allocation to a single digital servitization platform ecosystem at industry level | *Platform envelopment*<br>Unified platform with open APIs | *Data-driven business model*<br>Connection of Alpha ecosystem with other external ecosystems and selling raw data |

**Fig. 3.** Digital servitization (DS) evolutionary framework.

These theoretical insights also contribute to expanding the managerial and practical comprehension of the topics covered in this article. Specifically, the detailed case study provides valuable insights into how traditional manufacturers frame their digital servitization strategies and modify their BM dimensions, based on the deployment of AI platform-based services. Thus, the case study yields empirical evidence that can help managers of traditional manufacturing firms determine the appropriate course of action and strategic planning for a successful digital servitization transformation.

Overall, this study significantly enhances the theoretical and managerial understanding of B2B platform strategies in industrial markets (Stonig et al., 2022; Van Dyck et al., 2024), offering insights into the strategic, operational, and technological pivots that are reshaping the industry.

*5.2. Limitations and future research opportunities*

In addition to the contributions outlined, this study acknowledges certain limitations that present opportunities for future research endeavors.

Firstly, the reliance on a single case study poses limitations in terms of generalizability. Indeed, although the investigation of a single case company allows going for in-depth analysis of the main concepts, our findings, while insightful, may not be universally applicable across different contexts. Additional future qualitative research could extend and refine our analysis to various settings, assessing the validity of our understanding in different industrial sectors and contexts, thus evaluating the broader applicability of our findings.

Secondly, the study recognizes that the journey towards a platform BM is not linear. The utilization of digital technologies and AI platforms for servitization can vary significantly, as can the approaches to adapting BM components. Future research could employ a qualitative multiple case study methodology to compare and contrast the diverse strategies employed by manufacturers in digital servitization. Such comparative analyses would further enrich the literature by highlighting the spectrum of approaches in this field.

Third, the presented case study illustrates a successful transition of a multinational manufacturer towards digital servitization, which culminates in the provision of advanced digital services. Although these results represent important guidelines from both a managerial and academic point of view, the analysis and presentation of failed attempts to digital servitization and B2B platform-driven BM innovation also offer interesting learning resources for both scholars and industry professionals. Future studies may consider this as a promising opportunity to advance research on the topic.

Finally, there is a valuable opportunity for future research to adopt a quantitative approach. This would allow for the statistical testing of relationships among key variables, particularly examining the enabling role of AI platforms in the advancement and offering of digital services. Implementing such a methodology would provide empirical evidence to support or challenge the theoretical constructs proposed in this study, contributing to a more robust understanding of the dynamics at play in digital servitization.

**CRediT authorship contribution statement**

**Clara Filosa:** Writing – review & editing, Writing – original draft, Methodology, Investigation, Formal analysis, Data curation, Conceptualization. **Marin Jovanovic:** Writing – review & editing, Writing – original draft, Validation, Supervision, Project administration, Methodology, Investigation, Formal analysis, Data curation, Conceptualization. **Lara Agostini:** Writing – review & editing, Writing – original draft, Validation, Supervision, Project administration, Methodology, Investigation, Formal analysis, Data curation, Conceptualization. **Anna Nosella:** Writing – review & editing, Writing – original draft, Validation, Supervision, Project administration, Methodology, Investigation, Formal analysis, Data curation, Conceptualization.

**Acknowledgments**

The analysis of Alpha is based on interviews carried out with the company; we thank informants a lot for their availability. The conclusions and final interpretation are the output of the intellectual activity of the authors.

**Data availability**

The data that has been used is confidential.